\documentclass[letter,longauth,traditabstract]{aa} 
\usepackage{graphicx}
\usepackage{txfonts}
\begin{document}
    \title{The Aquila prestellar core population revealed by \emph{Herschel}
           \thanks{Herschel is an ESA space observatory with science instruments provided by 
           European-led Principal Investigator consortia and with important participation from NASA.} 
           \thanks{Figures 3, 4, 5, and 6 are only available in electronic format via http://edpsciences.org}}

   \subtitle{}

   \author{
          V. K\"onyves\inst{1}
          \and
            Ph. Andr\'e\inst{1}
          \and
            A. Men'shchikov\inst{1}
          \and
            N. Schneider\inst{1}
          \and
          D. Arzoumanian\inst{1}
          \and
          S. Bontemps\inst{1,2,3}
          \and
           M. Attard\inst{1}
          \and 
            F. Motte\inst{1}            
          \and
            P. Didelon\inst{1}
          \and
          A. Maury\inst{4} 
          \and
            and
            A. Abergel\inst{5}
          \and
          B. Ali\inst{6}
          \and
            J.-P. Baluteau\inst{7}
          \and 
            J.-Ph. Bernard\inst{8}
          \and 
            L. Cambr\'esy\inst{9}
          \and
            P. Cox\inst{10}
          \and
            J. Di Francesco\inst{11}
          \and
          A.M. di Giorgio\inst{12}
          \and
            M.J. Griffin\inst{13}
          \and 
            P. Hargrave\inst{13}
          \and 
            M. Huang\inst{14}
          \and
            J. Kirk\inst{13}
          \and
            J. Z. Li\inst{14} 
          \and
            P. Martin\inst{15} 
          \and
            V. Minier\inst{1}
          \and
            S. Molinari\inst{12}
          \and
            G. Olofsson\inst{16}
          \and
            S. Pezzuto\inst{12} 
          \and
            D. Russeil\inst{7} 
          \and 
            H. Roussel\inst{17} 
          \and
            P. Saraceno\inst{12}
          \and
            M. Sauvage\inst{1}
          \and
            B. Sibthorpe\inst{18}
          \and
            L. Spinoglio\inst{12}
          \and
            L. Testi\inst{4,19}
          \and 
            D. Ward-Thompson\inst{13}
          \and
            G. White\inst{20,21}
          \and
            C.D. Wilson\inst{22} 
          \and
            A. Woodcraft\inst{18}
          \and
            A. Zavagno\inst{7}
          }

   \institute{Laboratoire AIM, CEA/DSM--CNRS--Universit\'e Paris Diderot, IRFU/Service d'Astrophysique, C.E. Saclay,
              Orme des Merisiers, 91191 Gif-sur-Yvette, France
              \email{vera.konyves@cea.fr}
         \and
             CNRS/INSU, Laboratoire d'Astrophysique de Bordeaux, UMR 5804, BP 89, 33271 Floirac cedex, France
         \and
             Universit\'e de Bordeaux, OASU, Bordeaux, France
         \and 
             European Southern Observatory, Karl Schwarzschild Str. 2, 85748 Garching bei M\"unchen, Germany
         \and
             IAS, CNRS/INSU--Universit\'e Paris-Sud, 91435 Orsay, France
         \and
             Infrared Processing and Analysis Center, MC 100-22, California Institute of Technology, Pasadena, CA 91125, USA
         \and 
             Laboratoire d'Astrophysique de Marseille, CNRS/INSU--Universit\'e de Provence, 13388 
             Marseille cedex 13, France
         \and
             CESR \& UMR 5187 du CNRS/Universit\'e de Toulouse, BP 4346, 31028 Toulouse cedex 4, France
         \and           
             Observatoire Astronomique de Strasbourg, UMR 7550 CNRS/Universit\'e de Strasbourg, 11 rue de 
             l'Universit\'e, 67000, Strasbourg, France
         \and           
             IRAM, 300 rue de la Piscine, Domaine Universitaire, 38406 Saint Martin d'H\`{e}res, France         
         \and          
             National Research Council of Canada, Herzberg Institute of Astrophysics,
             University of Victoria, Department of Physics and Astronomy, Victoria, Canada         
         \and
             INAF-IFSI, Fosso del Cavaliere 100, 00133 Roma, Italy         
         \and  
             School of Physics and Astronomy, Cardiff University, Queens Buildings, The Parade, Cardiff CF24 3AA, UK
         \and           
             National Astronomical Observatories, Chinese Academy of Sciences, Beijing 100012, China
         \and
             Canadian Institute for Theoretical Astrophysics, University of Toronto, Toronto, ON M5S 3H8, Canada 
         \and
             Department of Astronomy, Stockholm Observatoty, AlbaNova University Center, Roslagstullsbacken 21, 
             10691 Stockholm, Sweden
         \and
             Institute d'Astrophysique de Paris, UMR 7095 CNRS, Universit\'e Pierre et Marie Curie, 
             98 bis Boulevard Arago, F-75014 Paris, France
         \and
             UK Astronomy Technology Centre, Royal Observatory Edinburgh, Blackford Hill, Edinburgh EH9 3HJ, UK
         \and
             INAF--Osservatorio Astrofisico di Arcetri, Largo Fermi 5, 50125 Firenze, Italy
         \and
             The Rutherford Appleton Laboratory, Chilton, Didcot OX11 0NL, UK
         \and
             Department of Physics and Astronomy, The Open University, Walton Hall, Milton Keynes MK7 6AA, UK
         \and
             Department of Physics and Astronomy, McMaster University, Hamilton, ON L8S 4M1, Canada         
          }

   \date{Received 1 April 2010; accepted 3 May 2010}

   \abstract{
   {The origin and possible universality of the stellar initial mass function (IMF) is a 
   major issue in astrophysics.
   One of the main objectives of the $Herschel$ Gould belt survey is to clarify 
   the link between the prestellar core mass function (CMF) and the IMF.
   We present and discuss the core mass function derived from $Herschel$ data for the large 
   population of prestellar cores discovered with SPIRE and PACS in the Aquila Rift cloud 
   complex at $d \sim260$~pc.
   We detect a total of 541 starless cores in the entire $\sim$11~deg$^2$ area of the field
   imaged at 70--500~$\mu$m with SPIRE/PACS.
   Most of these cores appear to be gravitationally bound, and thus prestellar in nature.      
   Our $Herschel$ results confirm that the shape of the prestellar CMF resembles the stellar 
   IMF, with much higher quality statistics than earlier submillimeter continuum ground-based surveys.} 

   \keywords{ISM: individual objects (Aquila Rift complex) -- Stars: formation 
             }}

   \maketitle
%

\section{Introduction}

The question of the origin and possible universality of the IMF, which is crucial for 
both star formation and galactic evolution, remains a major open problem in astrophysics. 
The $Herschel$ Space Observatory (Pilbratt et al. 2010), equipped with its two imaging 
cameras SPIRE (Griffin et al. 2010) and PACS (Poglitsch et al. 2010), provides a unique 
tool to address this fundamental issue. 
This problem and other key questions about the earliest phases of star formation and
evolution are central to the scientific motivation for the {\it Herschel} Gould belt 
survey, which will image the nearby ($d \leq 0.5$~kpc) molecular cloud complexes 
of the Gould belt using SPIRE at 250--500~$\mu$m and PACS at 70--160~$\mu$m 
(cf. Andr\'e \& Saraceno 2005 and Andr\'e et al. 2010).

Starting with the 1.2~mm continuum study of the Ophiuchus main cloud by Motte et al. (1998), 
several ground-based (sub)-millimeter dust continuum surveys of nearby, compact cluster-forming 
clouds such as $\rho$~Ophiuchi, Serpens, and Orion~B have uncovered 'complete' (but small) 
samples of prestellar cores whose associated core mass functions (CMF) resemble the stellar IMF 
(e.g., Motte et al. 1998; Johnstone et al. 2000; Enoch et al. 2006; Nutter \& Ward-Thompson 2007;
-- see also Alves et al. 2007 and Andr\' e et al. 2007).    
Albeit limited by small-number statistics at both ends of the CMF, these findings favor theoretical 
scenarios according to which the bulk of {\it the IMF of solar-type stars is largely determined 
by pre-collapse cloud fragmentation}, prior to the protostellar accretion phase 
(e.g., Hennebelle \& Chabrier 2008).
The problem of the origin of the IMF may thus largely reduce to a good understanding of the processes 
responsible for the formation and evolution of prestellar cores within molecular clouds. 
Apart from limited statistics, another limitation of ground-based submillimeter continuum determinations 
of the CMF is that they rely on uncertain assumptions about the dust properties (temperature and emissivity).
$Herschel$ is now making it possible to dramatically improve the quality of the statistics 
and to reduce the core mass uncertainties by performing direct measurements of the dust temperatures.

In this Letter, we use the initial results of the Gould belt survey, obtained toward the 
Aquila Rift complex as part of the science demonstration phase (SDP) of {\it Herschel},  
to present the first CMF derived from $Herschel$ data. We discuss the robustness 
of the resulting CMF based on several tests and simulations.
 
The Aquila molecular cloud complex is located at the high Galactic-longitude end of the Aquila Rift, 
in the neighbourhood of the Serpens star-forming region. Here we adopt a distance of 260~pc for the 
Aquila star-forming complex. For a detailed introduction of the region and a discussion of its distance, 
we refer the reader to Bontemps et al. (2010).  


   \begin{figure*}
   \begin{center}
 \begin{minipage}{1.0\linewidth}
    \resizebox{0.54\hsize}{!}{\includegraphics[angle=270]{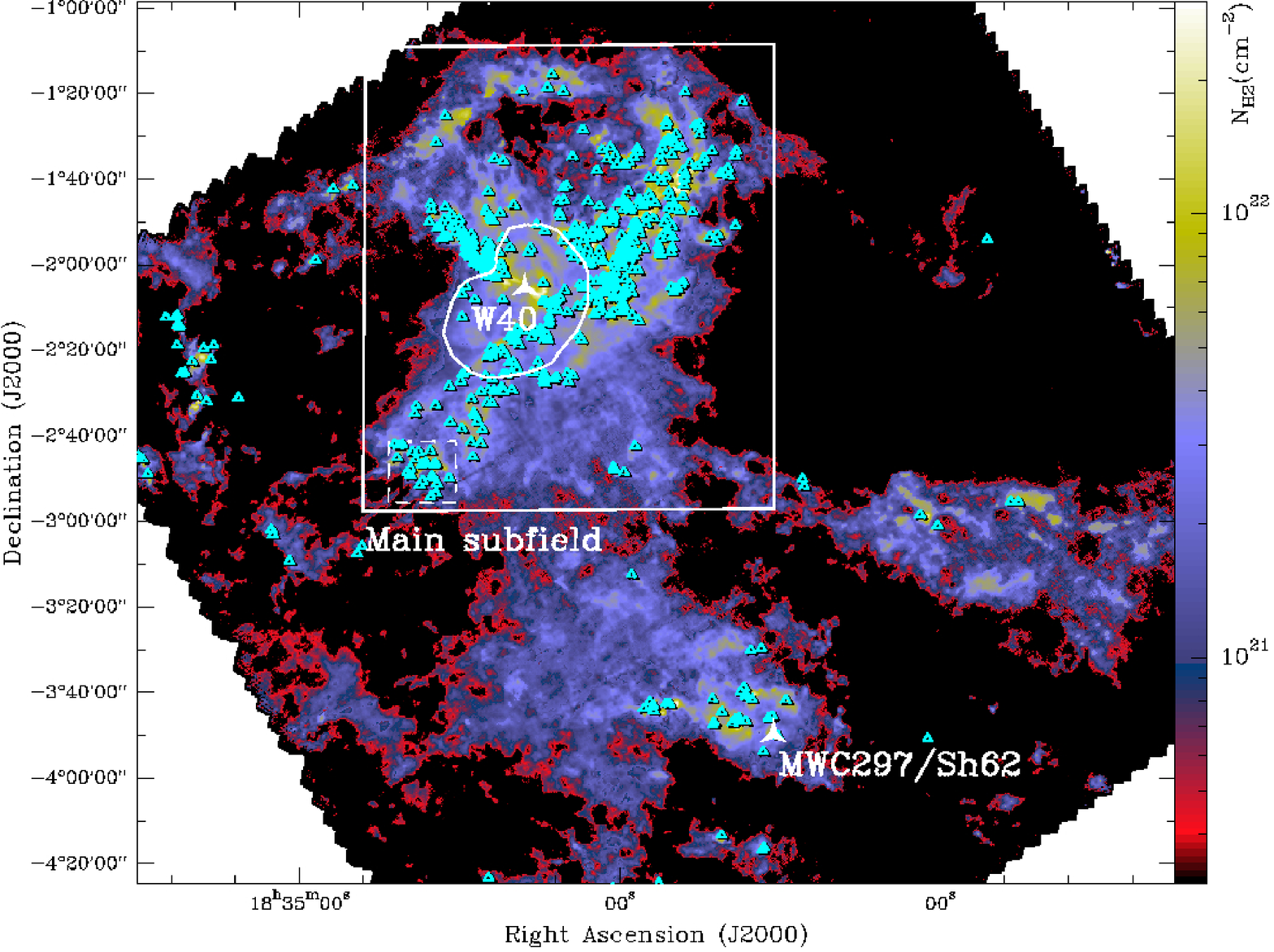}} 
    \hspace{2mm}
    \resizebox{0.44\hsize}{!}{\includegraphics[angle=270]{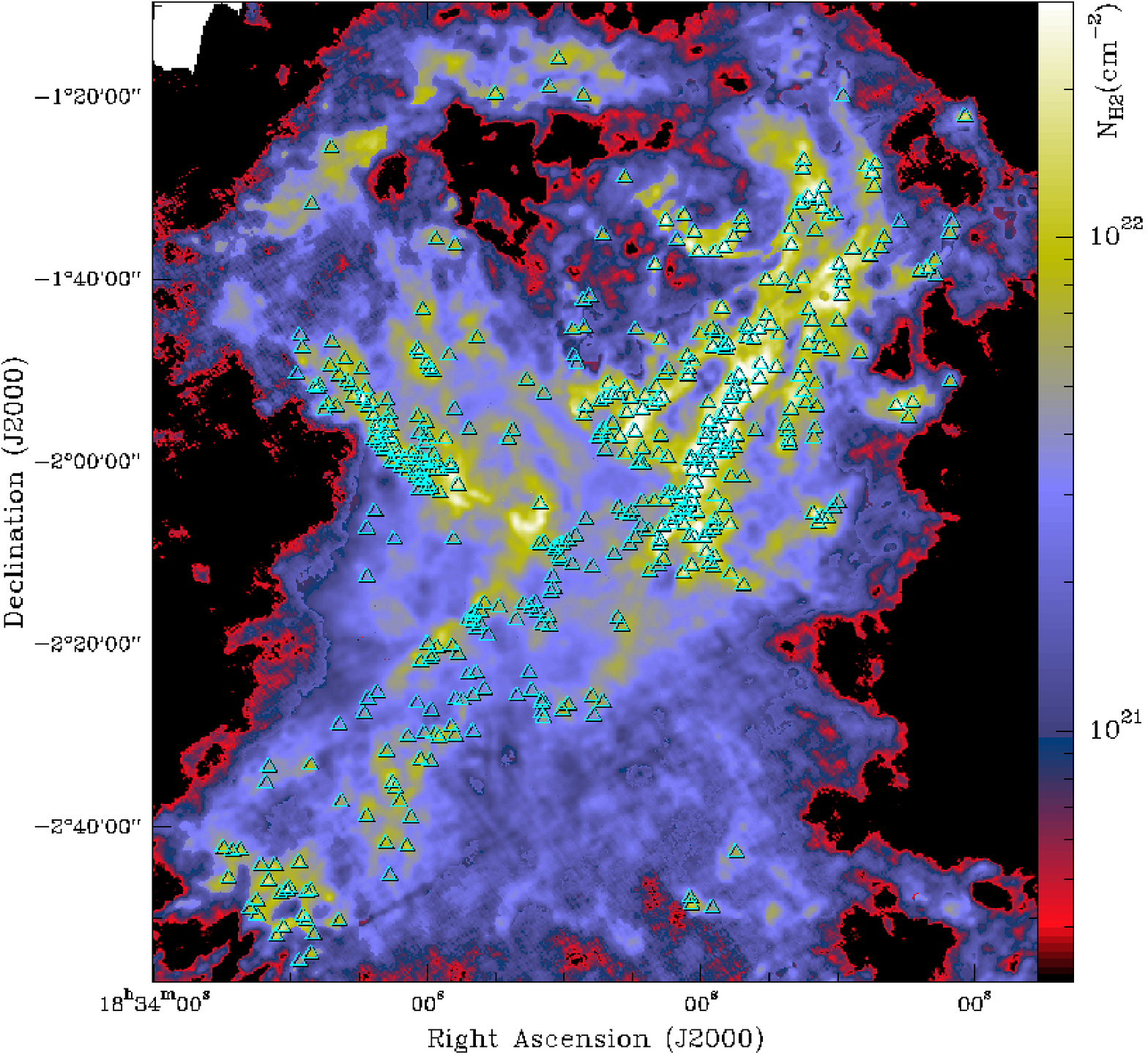}}
 \end{minipage}
   \end{center}
   \caption{{\bf(a)} Column density map derived from SPIRE/PACS observations of Aquila. 
            The subregion referred to as the main subfield in the text is marked by the white rectangle. The cyan 
            triangles mark the positions of the 541 starless cores identified in the entire field. 
            The locations of the HII regions W40 and MWC297/Sh62 are shown. The PDR region around W40 is framed by 
            white polygon, while the dashed square outlines the small region shown in more detail in online Fig.~\ref{Fig_zooms}a.  
            {\bf(b)} Same as (a) for the main subfield, with a total of 452 starless cores, marked by
            cyan triangles.} 
              \label{Fig_coldens}%
    \end{figure*}

\section{Observations and data reduction}

%
Our $Herschel$ observations of the Aquila Rift complex 
were taken on 24 October 2009 in the SPIRE/PACS parallel-mode. 
A common $\sim$11~deg$^2$ area was covered by  both SPIRE and PACS scan maps, with a 
scanning speed of 60$''$s$^{-1}$. The field was observed twice 
by performing cross-linked 
scans in two nearly orthogonal scan directions. The combination of the nominal and orthogonal 
coverages reduced the effects of 1/$f$ noise (see Sibthorpe et al. 2008).   
 
The PACS 
70~$\mu$m and 160~$\mu$m 
data were reduced with HIPE (Ardila et al. 2010) version 3.0.  
Standard steps of the default pipeline were applied 
starting from the raw data (level 0). 
We used file version 1 flat-fielding and responsivity in the calibration tree, instead of the built-in 
version 3 of those. Therefore, 
the flux density scale was corrected with the corresponding 
responsivity correction factors: 1.78 at 70~$\mu$m, and 1.43 at 160~$\mu$m.
Multi-resolution median transform (MMT) 
deglitching and second-order deglitching were also 
applied. 
Baselines were then 
subtracted by high-pass filtering, with a median filter width corresponding 
to the full length (180$'$--190$'$) of the scan legs taken from HSPOT (Frayer et al. 2007). The baseline 
fits were performed on the 'level 1' data using interpolation for the masked bright sources not to 
over-subtract true sky emission. In this way, we removed stripes and preserved spatial scales 
and diffuse emission up to the size of the maps. 
The final PACS maps were created using the photProject task, which performs simple projection
of the data cube on the map grid.

Our SPIRE observations at 250, 350, and 500~$\mu$m were reduced using
HIPE version 2.0 and the pipeline scripts delivered with this
version. These scripts were modified, e.g., to include observations taken
during the turnaround of the telescope. A median
baseline was applied to the maps and  the 'naive' map-making method was used. 
Online Fig.~\ref{Fig_spire} shows the 500/350/250~$\mu$m SPIRE images.
The PACS 160/70~$\mu$m images of the same field are shown in Bontemps et al. (2010).

For SPIRE, the absolute calibration uncertanty is estimated to be $\sim$15\%
(Griffin et al. 2010), while for PACS the absolute flux accuracy is within 10\% in the blue filter, and better 
than 20\% in the red filter (Poglitsch et al. 2010).  
The in-flight calibration of the SPIRE instrument is described by Swinyard et al. (2010).

Besides cross-correlating the SPIRE and PACS maps to test their relative astrometry, we compared 
the astrometry of the $Herschel$ images 
with publicly-available {\it Spitzer} 8~$\mu$m and 24~$\mu$m data, as well 
as high-positional accuracy ($<$1$''$) 3~mm IRAM Plateau de Bure observations of a small field at the 
center of the Aquila main filament (Maury et al. in prep.).
In this way, we corrected the {\it Herschel} images for small astrometric offsets ($\lesssim 6\arcsec$) remaining 
between the SPIRE and PACS maps, and achieved a final astrometric accuracy better 
than $\sim2\arcsec$.

   \begin{figure*}
   \centering
 \begin{minipage}{1.05\linewidth}
   \resizebox{0.445\hsize}{!}{\includegraphics[angle=270]{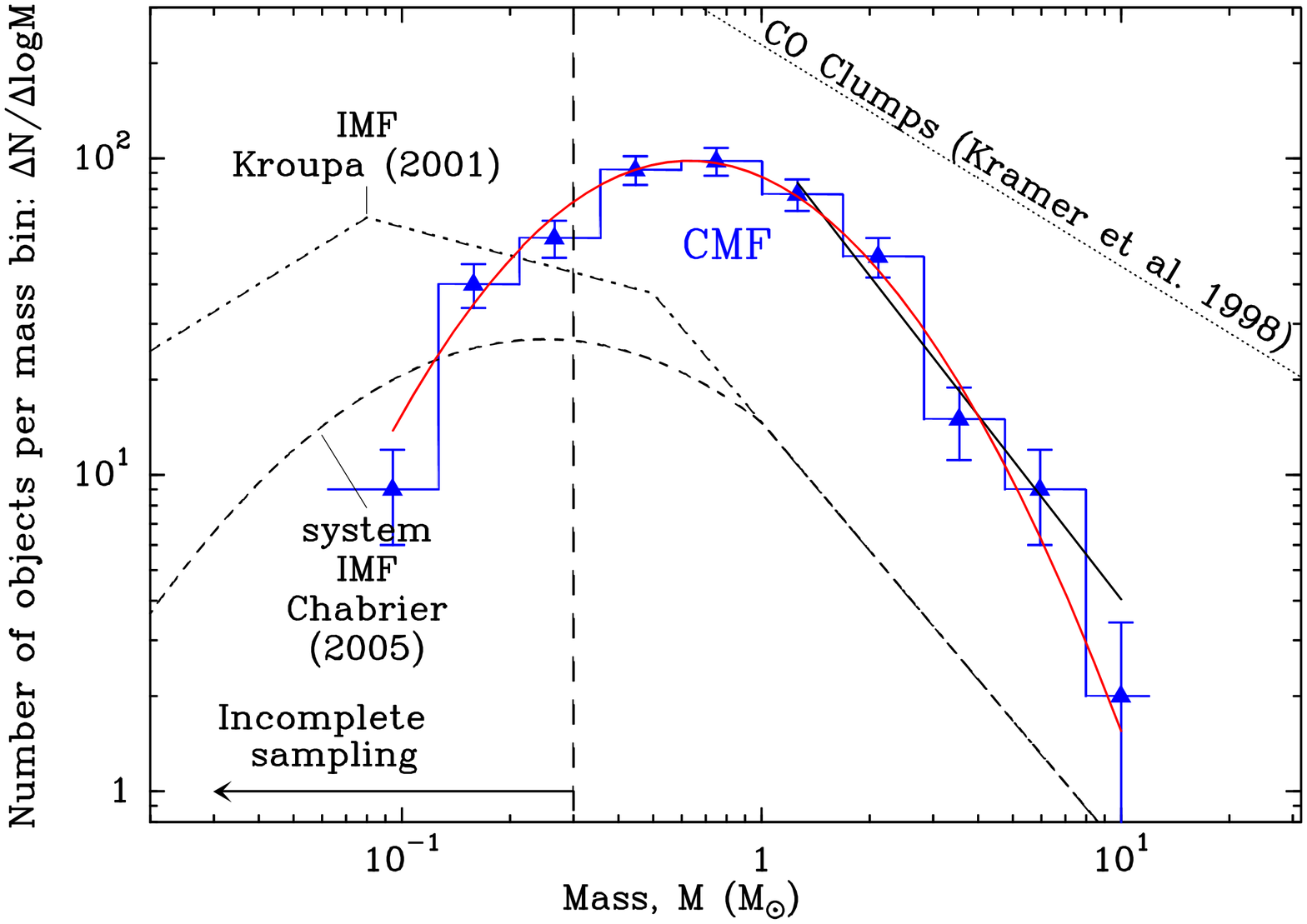}}  
   \hspace{10mm}
   \resizebox{0.445\hsize}{!}{\includegraphics[angle=270]{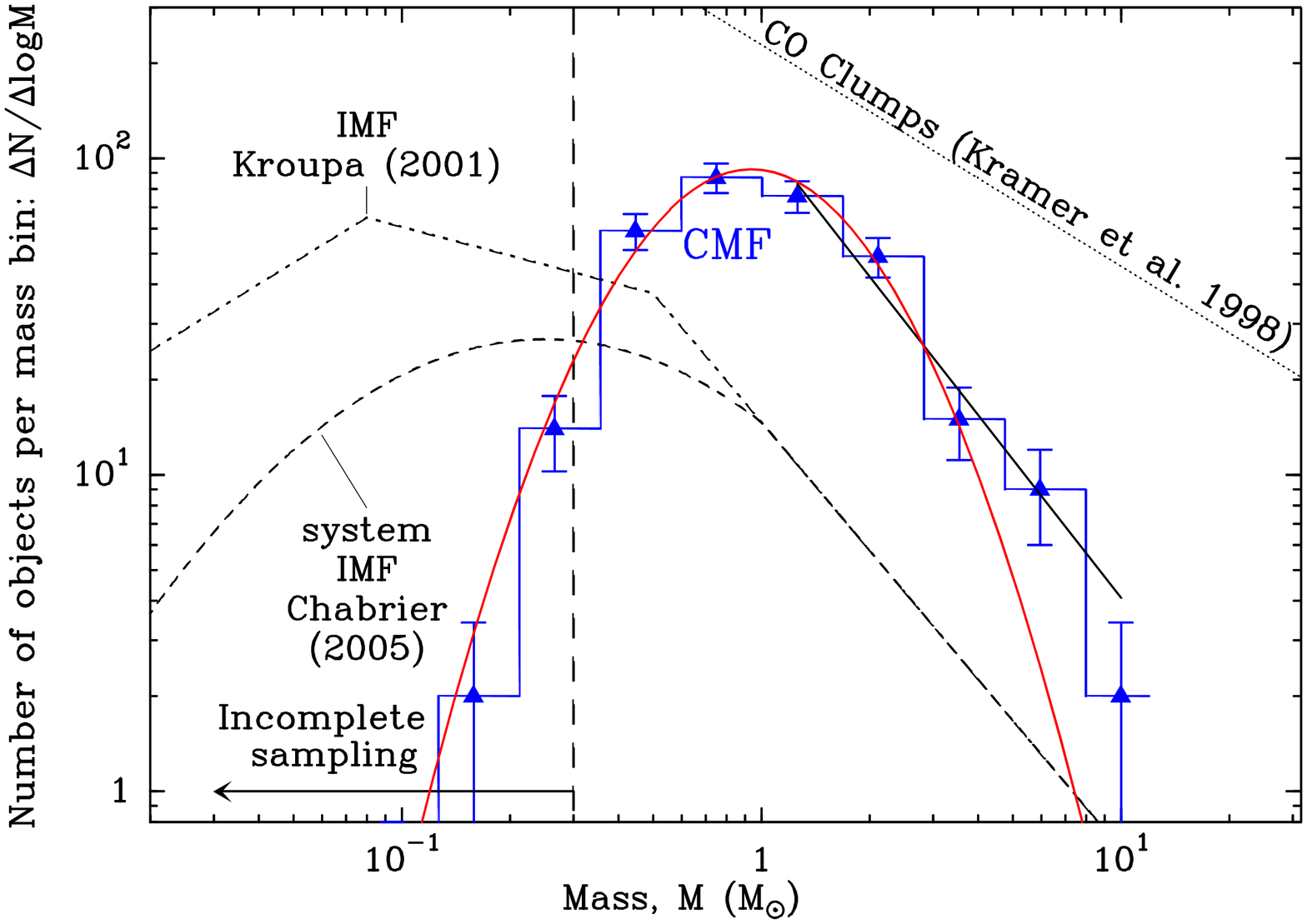}}
 \end{minipage}
   \caption{{\bf(a)} Differential mass function (d$N$/dlog$M$) of the 452 starless cores in the Aquila 
   main subfield, approximated with a lognormal fit (red curve). The error bars correspond to $\sqrt{N}$ 
   statistical uncertrainties. The core sample is estimated to be complete down to $\sim$0.2--0.3~$M_{\odot}$.
   The lognormal fit has a peak at $\sim$0.6~$M_{\odot}$ and a standard deviation of $\sim$0.42 in log$_{10}M$.  
   For comparison, the dash-dotted line shows the single-star IMF (e.g., Kroupa 2001), and the dashed 
   curve corresponds to the unresolved system IMF by Chabrier (2005).
   The dotted line shows a power law of the form d$N$/dlog$M$ $\propto$ $M^{-0.6}$, which is the typical 
   mass distribution of low-density CO clumps (see Kramer et al. 1998) 
   The high-mass end of the CMF is fitted by a power law (d$N$/dlog$M$ $\propto$ $M^{-1.5 \pm 0.2}$), 
   while the Salpeter IMF is d$N$/dlog$M$ $\propto$ $M^{-1.35}$. 
   {\bf(b)} Same as (a) for the subset of 314 candidate prestellar cores identified in the Aquila main subfield.
   Here, the best-fit power law to the high-mass end of the bound cores CMF gives the same result
   (d$N$/dlog$M$ $\propto$ $M^{-1.45 \pm 0.2}$), while the lognormal fit peaks at $\sim$0.9~$M_{\odot}$ and has a 
   standard deviation of $\sim$0.30. See text for discussion.   
}
              \label{Fig_mass_distr}%
    \end{figure*}

\section{Results and analysis}

Compact sources were extracted from the SPIRE/PACS images using {\textit{getsources}}, a multi-scale, multiwavelength 
source-finding algorithm briefly described in Men'shchikov et al. (2010).
Several sets of extractions were obtained, including one for the entire field and one for the main subfield 
of the Aquila complex (see Fig.~\ref{Fig_coldens}). 
At this early stage of the scientific exploitation of the $Herschel$ survey, we only considered robust sources 
with significant ($S/N > 7.5$) detections in at least two SPIRE bands, especially 
since the significance of the sources depends slightly on the adopted set of extractions.
  
For the Aquila main subfield (see  Fig.~\ref{Fig_coldens}), {\it Spitzer} 24~$\mu$m observations were used 
in combination with PACS 70~$\mu$m data to distinguish between starless cores and young (proto)stellar 
objects (YSOs).
In this subfield, objects detected in emission above the 5$\sigma$ level at 70~$\mu$m and/or 24~$\mu$m 
were classified as YSOs, while cores undetected in emission (or detected in absorption) at both 70~$\mu$m and 24~$\mu$m 
were classified as {\it starless}. This classification yielded 452 starless cores in the Aquila main subfield.

Outside the main subfield, we had to rely only on our PACS 70~$\mu$m data to distinguish 
between starless and protostellar cores. Based on the results obtained in the main subfield, we estimate 
that the lack of {\it Spitzer} 24~$\mu$m information leads only to a $\sim$3\% error in the classification.   
Altogether, we identified a total of 541 starless cores and 201 embedded YSOs in the entire field.   
The YSOs include $\sim$45--60 Class~0 protostars depending on the selection criteria (see Bontemps et al. 2010).

Based on our {\it Herschel} data, we constructed dust temperature ($T_{\rm d}$) and column density ($\Sigma$) 
maps. To do this, we first smoothed all {\it Herschel} images to the 500~$\mu$m resolution (36.9$''$) 
and reprojected them to the same 6$''$ pixel grid. Weighted spectral energy distributions (SEDs) were 
then constructed for all map pixels from the 5 observed SPIRE/PACS wavelengths. 

Assuming single-temperature dust emission, we fitted each SED by a grey-body function of the form
$I_{\nu}$ = $B_{\nu}(T_{\rm d})(1-e^{-\tau_{\nu}})$, where $I_{\nu}$ is the observed surface brightness 
at frequency $\nu $, $\tau_{\nu} = \kappa_{\nu} \Sigma$ is the dust optical depth, and $\kappa_{\nu} $ is 
the dust opacity per unit (dust$+$gas) mass, which was approximated by the power law 
$\kappa_{\nu} = 0.1~(\nu/1000~{\rm GHz})^{\beta}$ cm$^2$/g (cf. Beckwith et al. 1990). 
The dust 
emissivity index $\beta$ was fixed to 2 (e.g., Hildebrand 1983).

Each SED data point was weighted by 1/$\sigma^2$, where the rms noise $\sigma$ was estimated in an emission-free 
region of the map at the relevant wavelength, and the calibration uncertainties were also included. The two free 
parameters $T_{\rm d}$ and $\Sigma$ were derived from the grey-body fit to the 5 $Herschel$ data points for all pixels 
for which the fit was successful. 
Map pixels for which the fit was unsuccessful or unreliable were assigned the median dust temperature 
of the successful fits. Likewise, the column density along the line of sight to pixels with unreliable fits was 
estimated directly 
from the surface brightness measured at the longest wavelength with a reliable detection, assuming the median 
dust temperature of the successful fits.  
 
A similar SED fitting procedure was employed to estimate the dust temperature, 
column density, and mass of each core.
Here, the SEDs were constructed from the integrated flux densities measured by {\textit{getsources}} for 
the extracted sources. 
Ignoring the distance uncertainty (see discussion in Appendix A of Andr\'e et al. 2010), the core mass uncertainty 
is typically a factor of $\sim2$, mainly due to uncertainties in the dust opacity law ($\kappa_\nu$).

Monte Carlo simulations were carried out to estimate the completeness level of our SPIRE/PACS survey.
We first constructed clean maps of the background emission at all 
$Herschel$ wavelengths by subtracting the emission of the compact sources identified with {\textit{getsources}} 
from the SPIRE/PACS images of Aquila. 
We then inserted a population of $\sim700$ model starless cores and $\sim200$ model protostars at random positions 
in the clean-background images to generate a full set of synthetic $Herschel$ images of the Aquila region.
The model cores were given a realistic mass distribution in the 0.01$-$10~$M_{\odot}$ range 
and were assumed to follow a $M \propto R$ mass versus size relation.
The emission from the synthetic cores
was based on spherical 
dust radiative transfer models (Men'shchikov et al. in prep.). 
Compact source extraction of several sets of these synthetic skies was performed with {\textit{getsources}} 
in the same way as for the observed images.
Based on these simulations, we estimate that our $Herschel$ census of prestellar cores 
is 75\% and 85\% complete above a core mass of $\sim$0.2 and $\sim$0.3 M$_{\odot}$, respectively, 
in most of the field. Likewise, our census of embedded protostars is more than 90\% complete down to 
$L_{\rm{bol}} \sim 0.2~L_{\odot}$. Our survey may however be less complete than these values in the 
high background region around the W40 PDR (see white polygon in Fig.~\ref{Fig_coldens}a).

\section{Discussion and conclusions}

\subsection{Prestellar nature of the Aquila starless cores}

In this paper, we follow the naming convention that a dense core is called {\it prestellar} if it 
is starless {\it and} gravitationally bound (cf. Andr\'e et al. 2000, Di Francesco et al. 2007).
In other words, prestellar cores represent the subset of starless cores that are most likely to 
form (proto)stars in the future.

Strictly speaking, spectroscopic observations would be required to derive virial masses for 
the cores and determine whether they are gravitationally bound or not. 
However, millimeter line observations in dense gas tracers such as 
N$_2$H$^+$ show that thermal motions generally dominate over non-thermal motions in 
starless cores (e.g., Andr\'e et al. 2007). Assuming that this is indeed the case for the Aquila cores 
observed here, we may use the critical Bonnor-Ebert (BE) mass, 
$M_{{\rm BE}}^{{\rm crit}}~\approx~2.4~R_{{\rm BE}}~a^2 / G$,  as a surrogate 
for the virial mass, where $R_{{\rm BE}}$ is the BE radius, 
$a$ is the isothermal sound speed, and $G$ is the gravitational constant.  
The critical BE mass may also be expressed as $M_{{\rm BE}}^{{\rm crit}} \approx 1.18 {a^4 \over G^{3/2}} P_{{\rm ext}}^{-1/2} $, 
where $P_{{\rm ext}}$ is the external pressure, which may be estimated as a function of the column density 
of the local background cloud, $\Sigma_{cl}$, as $P_{{\rm ext}} \approx 0.88~G~\Sigma_{cl}^2$ (McKee \& Tan 2003).
For each object, we derived two estimates of the BE mass: (1) $M_{BE}(R_{obs})$ as a function of the observed core radius
assuming a gas temperature of 10~K, and (2) $M_{BE}(\Sigma_{cl})$ as a function of the local background column density 
measured from both our source-subtracted $Herschel$ images and the near-IR extinction map of 
Bontemps et al. (2010 -- see also Fig.~\ref{Fig_coldens_compare}b). 
We then calculated BE mass ratios of $\alpha_{\rm BE} \equiv \rm{max}[M_{BE}(R_{obs}), M_{BE}(\Sigma_{cl})] / M_{obs} $ 
and selected candidate prestellar cores to be the subset of starless cores for which $\alpha_{\rm BE}  \lesssim 2$. 
Based on this criterion, 314 (or $\sim69\% $) of the 452 starless cores of the main subfield 
and 341 (or $\sim63\% $)  of the 541 starless cores of the entire field 
were found to be bound and classified as good candidate prestellar cores.
These high fractions of bound objects are consistent with the locations of 
the Aquila starless cores in a mass versus size diagram 
(see online Fig.~\ref{Fig_mass_size} for the main subfield and Fig.~4 of Andr\'e et al. 2010 
for the entire field).

The self-gravitating character of most $Herschel$ cores in Aquila is supported further by an independent 
analysis of their internal column density contrasts. We here define the internal column density 
contrast of a core as $\Sigma_{peak}/$$\bar{\Sigma}_{core}$, where $\Sigma_{peak}$ and $\bar{\Sigma}_{core}$
are the peak and mean column densities of the core, respectively. Assuming optically thin dust emission 
at $Herschel$ wavelengths and neglecting any dust temperature/opacity gradient, the internal column 
density contrast can be estimated from the core intensity values in the same form, as
$I_{\nu}^{\rm peak}/\bar{I}_{\nu}$. 
The internal column density contrast is closely related to the degree of central concentration of a core
defined by Johnstone et al. (2000) as $C =1-(\bar{\Sigma}_{core}/\Sigma_{peak})$. 
Based on their radial intensity profiles (cf. online Fig.~\ref{Fig_zooms}b), the Aquila starless cores have a median 
internal column density contrast $\sim4$. 
For comparison, the internal column density contrast is larger than 3.6 for supercritical self-gravitating BE spheres,
while it is only 1.5 for a uniform-density, non-self-gravitating object (Johnstone et al. 2000). 

As a final check, we also estimated the typical column density contrast of the $Herschel$ cores over the
local background. For a critically self-gravitating BE core, the mean column density 
$\bar{\Sigma}_{\rm BE}^{{\rm crit}}~\approx~1.56~(P_{{\rm ext}}~/ G)^{1/2}$
is expected to exceed the column density of the local background cloud $\Sigma_{cl}$ by a factor of 1.46, 
if $P_{{\rm ext}}~\approx~0.88~G~\Sigma_{cl}^2$ (McKee \& Tan 2003). 
The candidate prestellar cores identified with $Herschel$ have a median column density contrast of  $\sim1.5$ 
over the local background, which is consistent with the conclusion that they are self-gravitating.
%

\subsection{Prestellar core mass function in Aquila}

Figure~\ref{Fig_mass_distr}a shows the mass distribution (CMF) of the 452 starless cores identified in 
the Aquila main subfield.
A  power-law fit to the high-mass end of this CMF gives d$N$/dlog$M$ $\propto$ $M^{-1.5 \pm 0.2}$
for $M_{\rm core} > 2~M_\odot$, which is very close to the Salpeter power-law IMF
(d$N/$dlog$M$ $\propto$ $M^{-1.35}$ -- Salpeter 1955). 
The CMF of the 541 starless cores identified in the entire field, shown by Andr\'e et al. (2010), 
has a very similar shape.

The robustness of the derived CMF was tested by selecting several subsets of sources, based e.g.,
on their physical nature or locations in the SPIRE/PACS maps.
In particular, we compared the mass spectrum of starless and prestellar cores. 
Figure~\ref{Fig_mass_distr}a was then compared with Fig.~\ref{Fig_mass_distr}b, which shows the 
differential mass function of 314 bound cores in the main subfield. We obtained the same best-fit power law
fit to the high-mass end (d$N$/dlog$M$ $\propto$ $M^{-1.45 \pm 0.2}$) as for Fig.~\ref{Fig_mass_distr}a.
Only the low-mass end changed when selecting the subset of candidate prestellar cores. 

We also selected subsets of starless cores based on their locations in the Aquila field.  
Along lines of sight to the HII region W40, the associated photon-dominated region (PDR, Shuping et al. 1999) 
is the source of very strong extended background emission at all infrared wavelengths.
The strong background emission in the {\it Spitzer} 24~$\mu$m and PACS 70~$\mu$m images 
makes it more difficult to discriminate between YSOs and compact starless cores, 
implying that our census of starless cores is less reliable in this area.
We thus constructed another version of the Aquila main subfield CMF, excluding the 83 cores identified 
toward the PDR region. 
The excluded region of unusually high infrared background emission (see in Fig.~\ref{Fig_coldens}a) 
was defined using the dust temperature map shown in Bontemps et al. (2010).
The high-mass end of this CMF can be fitted by a very similar power law to that of Fig.~\ref{Fig_mass_distr}a: 
d$N$/dlog$M$ $\propto$ $M^{-1.5 \pm 0.3}$.

Thanks to the large number of starless cores identified with $Herschel$ in Aquila
(541 cores in the entire field), we have been able to consider several core subsamples and construct 
a statistically meaningful CMF in each case. We confirm that the shape of the prestellar 
CMF resembles the stellar IMF, using data of far higher quality statistics than earlier submillimeter 
ground-based surveys and more accurate core masses. Based on simulations, we conclude 
that our mass distributions are robust and do not depend strongly on 
different sets of extracted sources.

The column density maps shown in Fig.~\ref{Fig_coldens} and online Fig.~5 illustrate the tight  
correlation between the spatial distribution of the prestellar cores and the filamentary structure 
revealed by the $Herschel$ images (Men'shchikov et al. 2010). 
The difference between our two SDP fields (Aquila Rift and Polaris Flare), in terms of filamentary 
structure and core mass distribution, is highlighted and discussed in Andr\'e et al. (2010).

\begin{acknowledgements}
PACS has been developed by a consortium of institutes led by MPE
(Germany) and including UVIE (Austria); KUL, CSL, IMEC (Belgium); CEA,
OAMP (France); MPIA (Germany); IFSI, OAP/AOT, OAA/CAISMI, LENS, SISSA
(Italy); IAC (Spain). This development has been supported by the funding
agencies BMVIT (Austria), ESA-PRODEX (Belgium), CEA/CNES (France),
DLR (Germany), ASI (Italy), and CICT/MCT (Spain).

SPIRE has been developed by a consortium of institutes led by
Cardiff Univ. (UK) and including Univ. Lethbridge (Canada);
NAOC (China); CEA, LAM (France); IFSI, Univ. Padua (Italy);
IAC (Spain); Stockholm Observatory (Sweden); Imperial College
London, RAL, UCL-MSSL, UKATC, Univ. Sussex (UK); Caltech, JPL,
NHSC, Univ. Colorado (USA). This development has been supported
by national funding agencies: CSA (Canada); NAOC (China); CEA,
CNES, CNRS (France); ASI (Italy); MCINN (Spain); SNSB (Sweden);
STFC (UK); and NASA (USA).  

V. K. acknowledges support from the "CONSTELLATION" EC FP6 Marie Curie Research 
Training Network (MRTN-CT-2006-035890).

\end{acknowledgements}

\onlfig{3}{
   \begin{figure*}
    \begin{minipage}{1\linewidth}
     \resizebox{0.55\hsize}{!}{\includegraphics[angle=270]{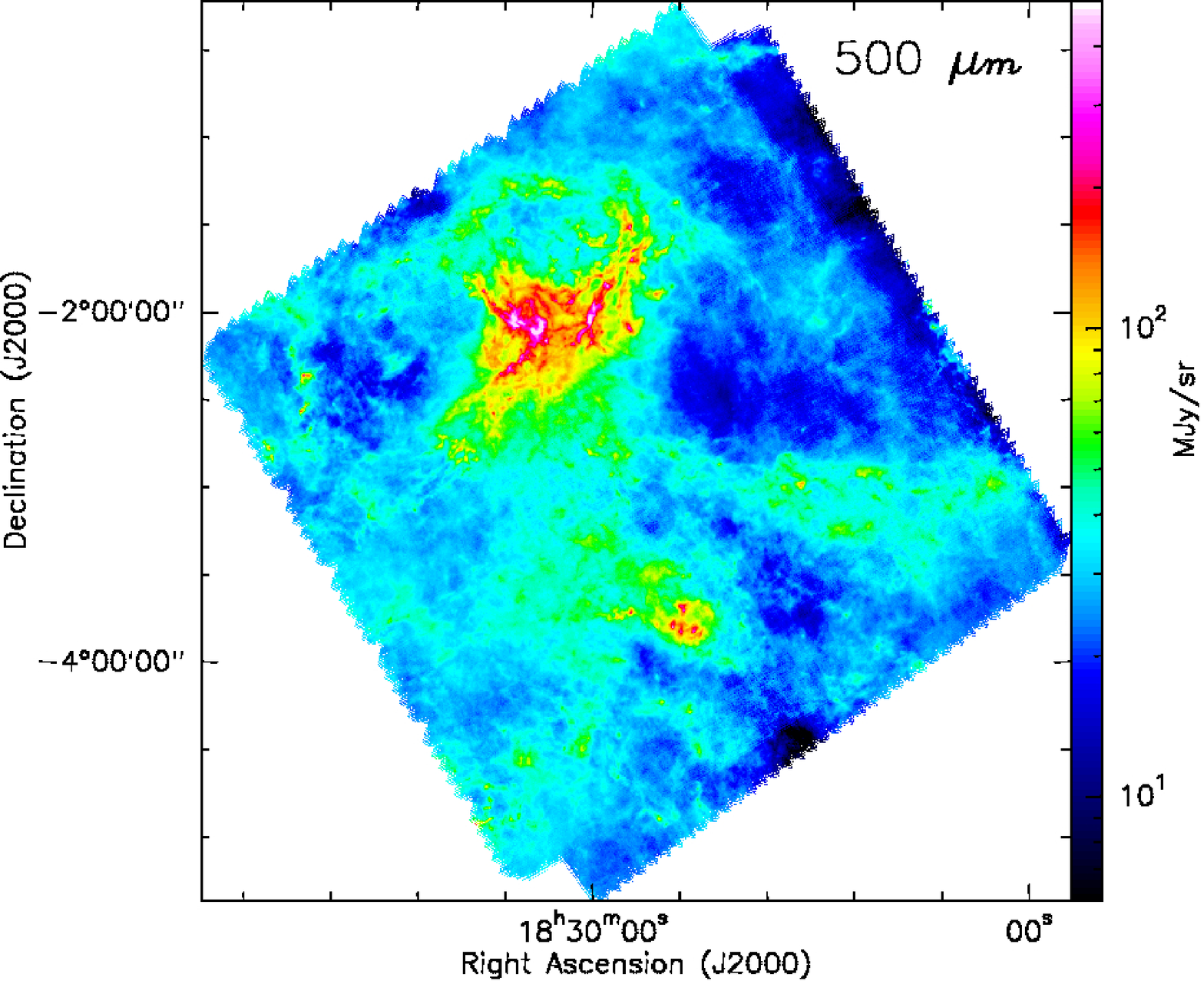}}
       \end{minipage}   

     \begin{minipage}{1\linewidth}
     \resizebox{0.55\hsize}{!}{\includegraphics[angle=270]{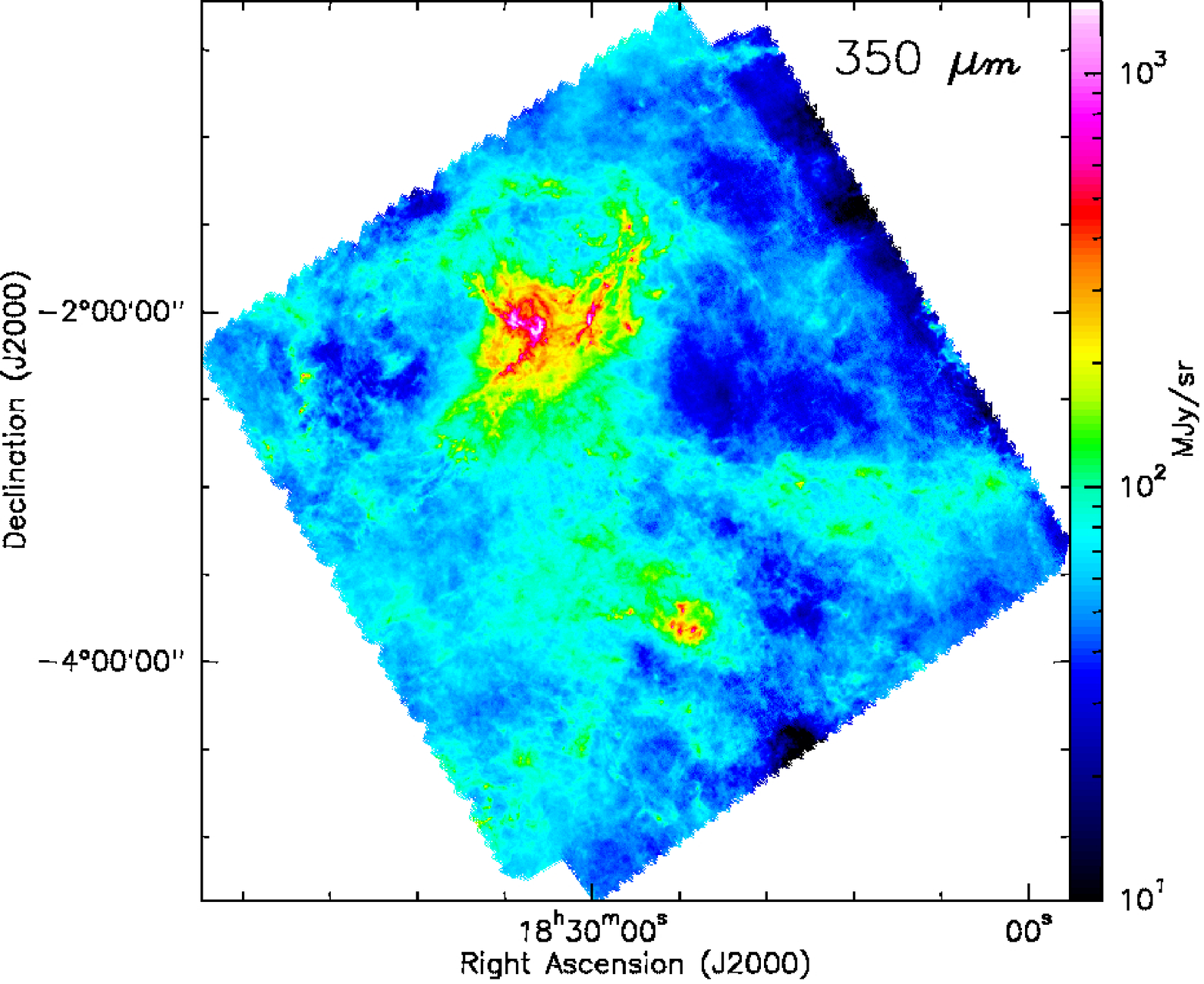}}
     \end{minipage}

     \begin{minipage}{1\linewidth}
     \resizebox{0.55\hsize}{!}{\includegraphics[angle=270]{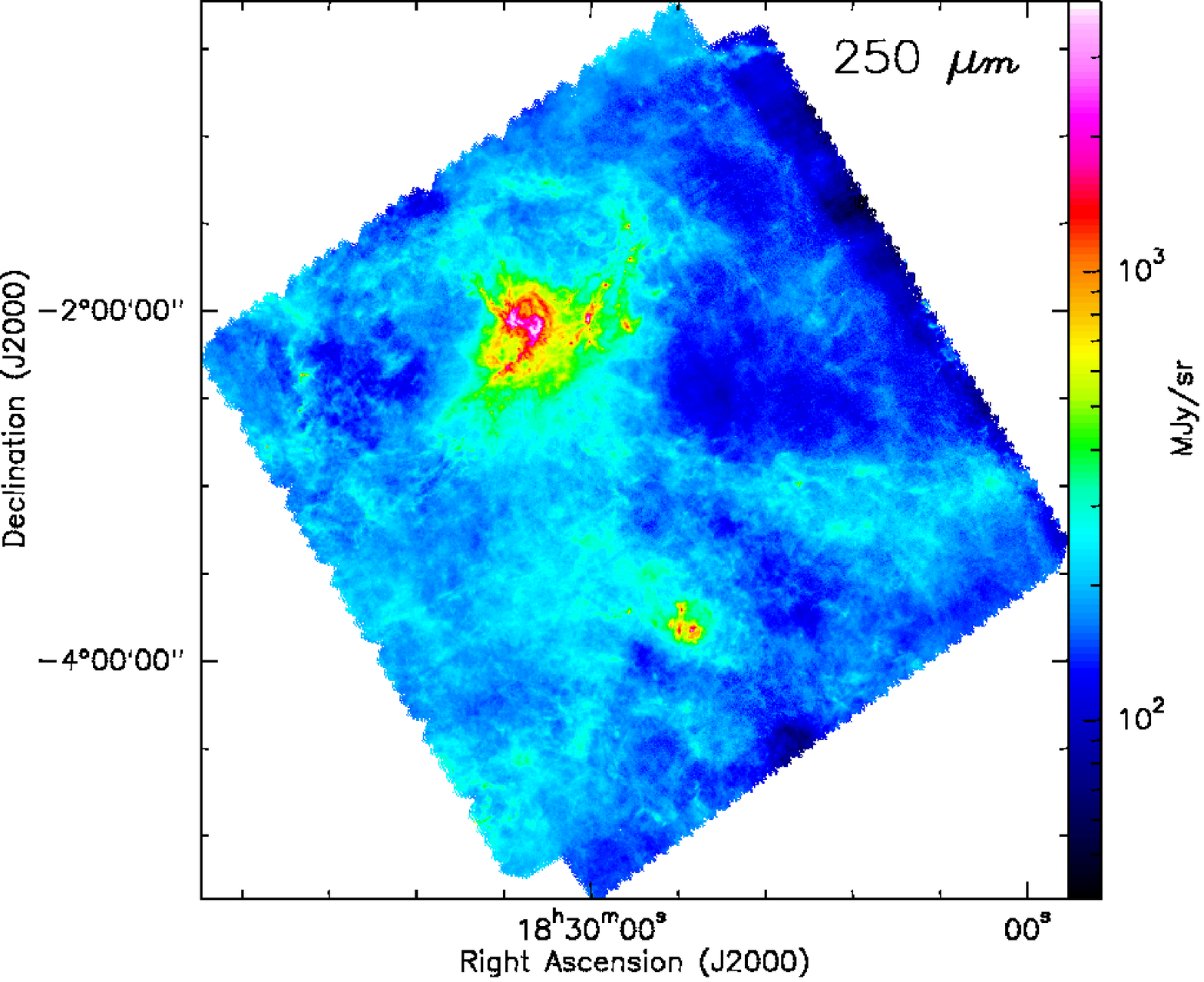}}
     \end{minipage}   
   \caption{SPIRE 500~$\mu$m (top), 350~$\mu$m (middle), and 250~$\mu$m (bottom) images of the 
            Aquila SDP field. See details about data reduction and map making in Section 2. 
            The corresponding PACS 160~$\mu$m and 70~$\mu$m images are shown in Bontemps et al. (2010). }
              \label{Fig_spire}%
    \end{figure*}
}

\onlfig{4}{
   \begin{figure*}[!!!t]   
 \centering
 \begin{minipage}{0.5\linewidth}
      \centerline{\resizebox{1.0\hsize}{!}{
      \includegraphics[angle=270]{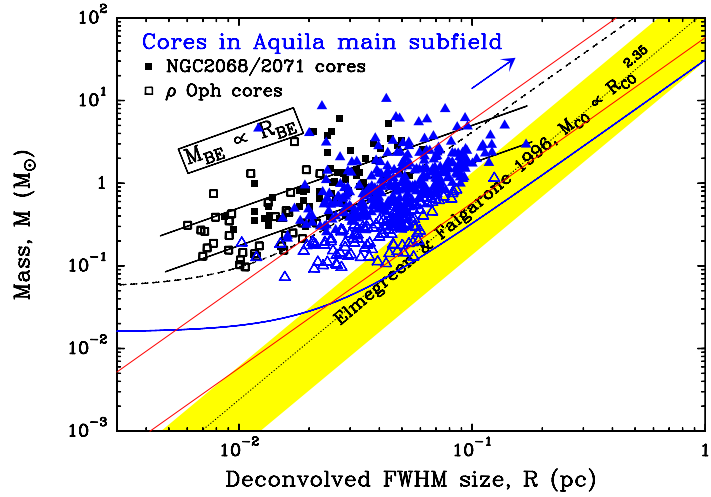}}}
 \end{minipage}
      \caption{Mass versus size diagram comparing the locations of the 452 starless cores identified with $Herschel$ in the 
               Aquila main subfield to both models of critical isothermal Bonnor-Ebert spheres at $T = 7$~K and $T = 20$~K 
               (black solid lines) and observed prestellar cores from the ground-based 
               (sub)-mm continuum studies of $\rho$ Ophiuchi and NGC2068/2071 by e.g., Motte et al. (1998).
               The masses of the $Herschel$ cores were derived as explained in the text and their sizes measured at 250~$\mu$m. 
               The 314 candidate prestellar cores of the main subfield
               are shown as filled triangles, while the other starless cores are shown as open triangles.
               The mass--size correlation 
               observed for diffuse CO clumps is also displayed (shaded yellow band $-$ Elmegreen \& Falgarone 1996). 
               The  typical (5$\sigma$) detection 
               threshold of current ground-based (sub)mm (e.g., MAMBO, SCUBA) surveys  at $d = 150$~pc 
               is shown by the dashed curve. The estimated $5\sigma$ detection threshold of our SPIRE 250~$\mu$m observations 
               is shown by the blue curve.
               The lower and upper red diagonal lines correspond to constant 10$^{21}$~cm$^{-2}$ and 
               10$^{22}$~cm$^{-2}$ mean column densities, respectively. The arrow indicates the global shift of 
               the sources if a distance of 400~pc were adopted instead of 260~pc for the Aquila Rift complex 
               (see also Appendix A of Andr\'e et al. 2010).}
         \label{Fig_mass_size}
   \end{figure*}
}

\onlfig{5}{
   \begin{figure*}[!!!b]
   \centering
 \begin{minipage}{1.1\linewidth}
   \begin{minipage}{0.424\linewidth}
   \resizebox{1\hsize}{!}{\includegraphics[angle=270]{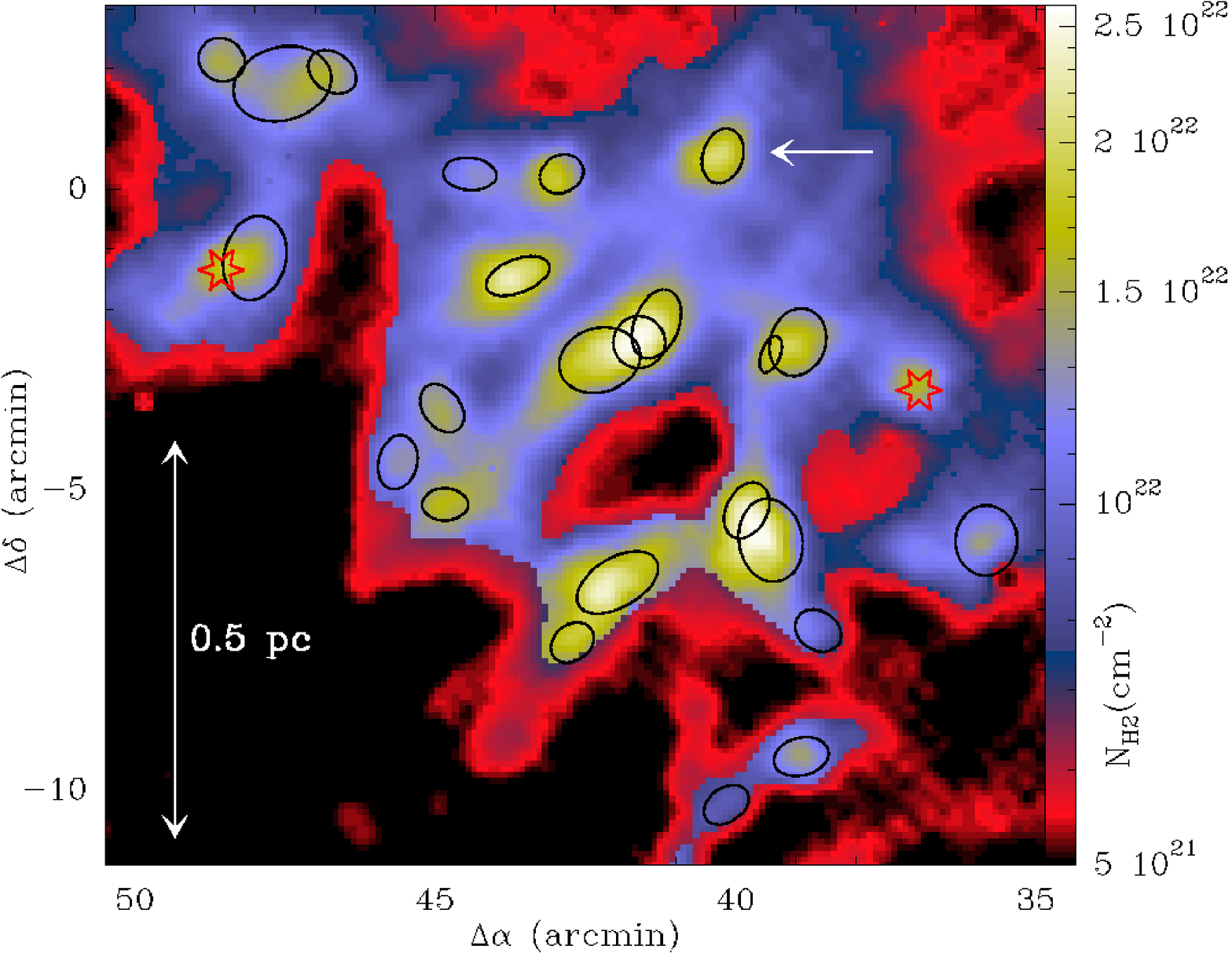}}
   \end{minipage}
   \hspace{2mm}
   \begin{minipage}{0.466\linewidth}
   \vspace{2mm}
   \resizebox{1\hsize}{!}{\includegraphics[angle=90]{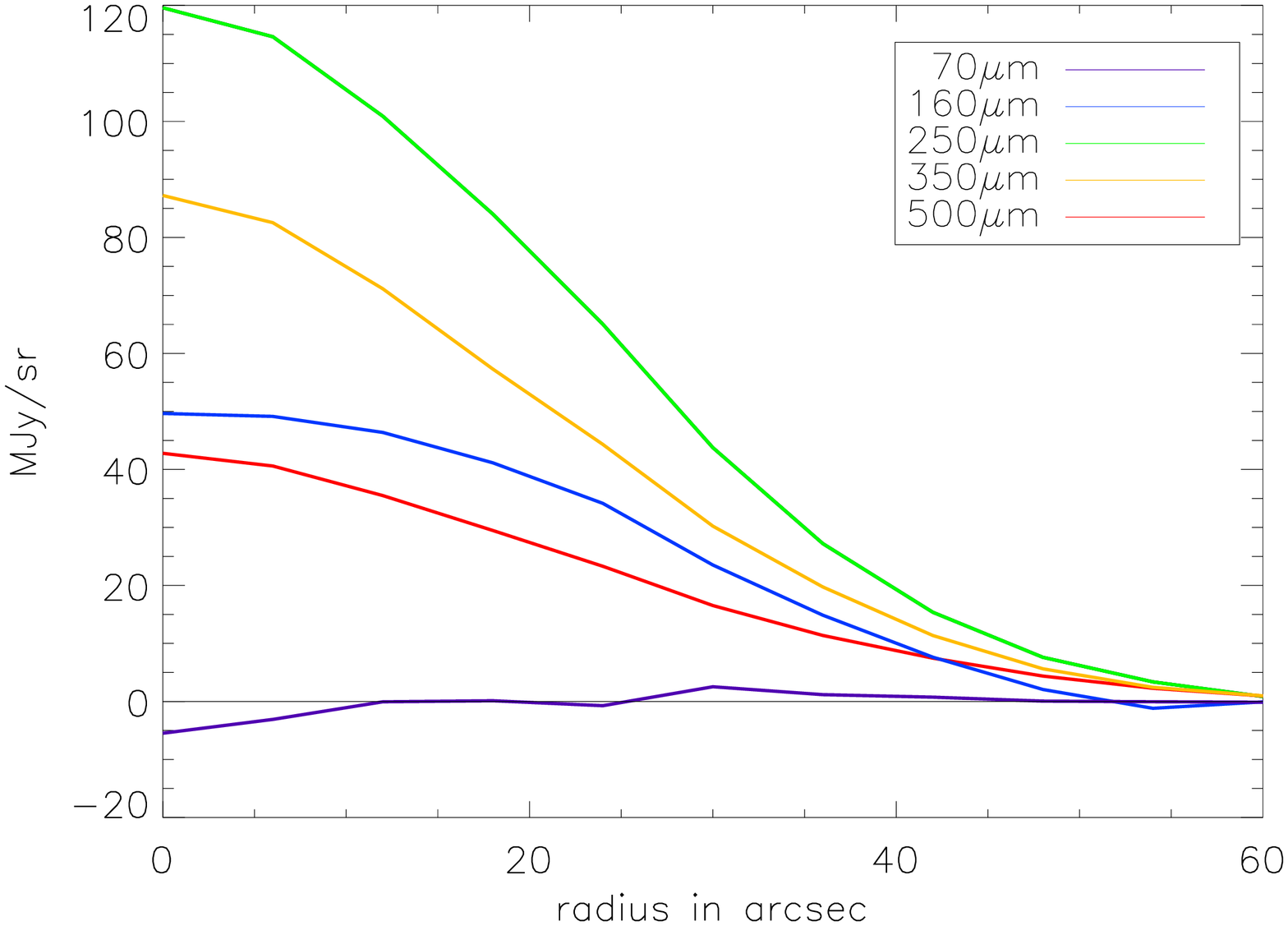}}

   \end{minipage}
 \end{minipage}
   \caption{{{\bf(a)} Close-up column density image (selected from Fig.~\ref{Fig_coldens}a) showing a close view
             of several starless cores identified with {\textit{getsources}}. The black ellipses mark the major and minor 
             FWHM sizes determined by {\it getsources} for these cores at $\lambda = 250\, \mu$m. Two protostars are also shown by red
             stars. {\bf(b)} Intensity profiles returned by {\it getsources} for the 
             starless core marked by the arrow in Fig.~\ref{Fig_zooms}a, at the five $Herschel$ wavelengths.}}
              \label{Fig_zooms}%
    \end{figure*}
}

\onlfig{6}{
   \begin{figure*}[!!!b]
   \centering
 \begin{minipage}{1.1\linewidth}
   \begin{minipage}{0.445\linewidth}
   \resizebox{1\hsize}{!}{\includegraphics[angle=270]{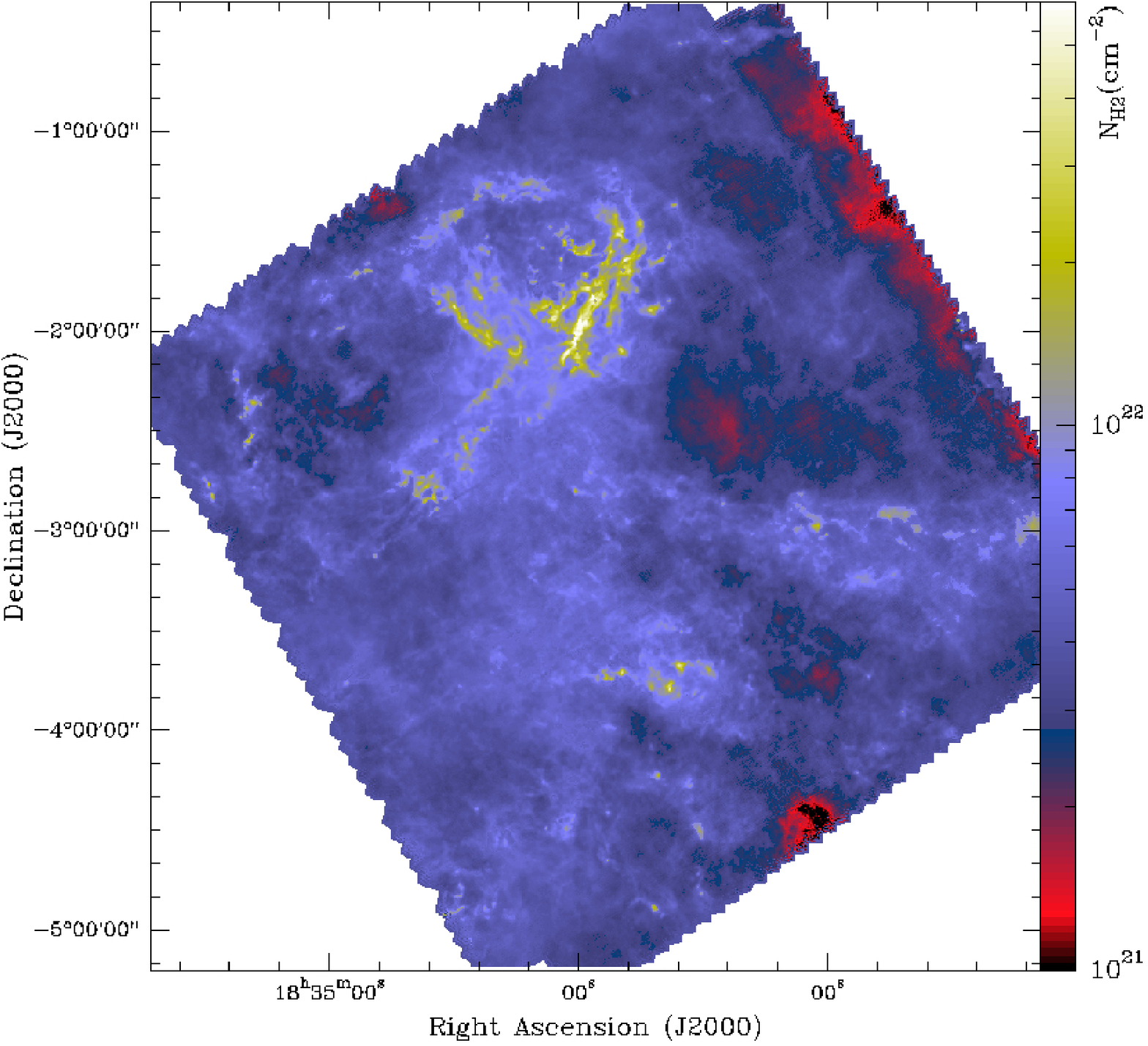}}
   \end{minipage}
   \hspace{2mm}
   \begin{minipage}{0.445\linewidth}
   \resizebox{1\hsize}{!}{\includegraphics[angle=270]{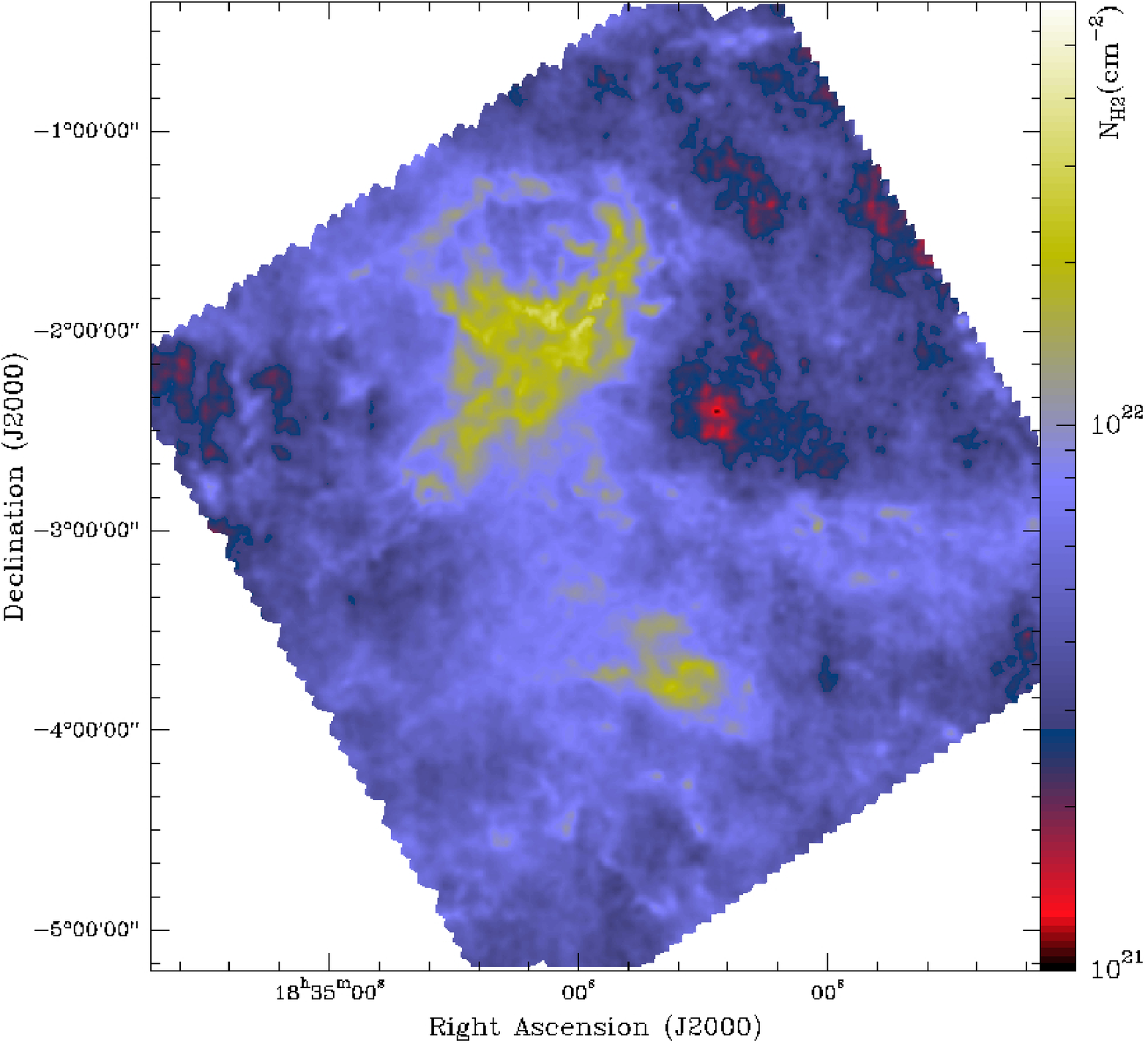}}
   \end{minipage}
 \end{minipage}
   \caption{{{\bf(a)} Column density map of the Aquila entire field derived from $Herschel$ data. 
   The effective FWHM resolution is 36.9$''$.  
   Unlike in Fig.~\ref{Fig_coldens}, a uniform offset  $N_{{\rm H_{2}}}^{\rm off} =  3.8 \times 10^{21}\, \rm{cm}^{-2}$ 
   has been added in order to optimize the match with the near-IR extinction  shown in (b).
   (Indeed, the $Herschel$ mapping does not constrain the zero level of the background emission.)
   {\bf(b)} Near-IR extinction map of the same field based on 2MASS data (see Bontemps et al. 2010) and expressed in units of column density, 
   using the relation $N_{{\rm H_{2}}} = 10^{21} \, \rm{cm}^{-2} \times A_V$. 
   The resolution is 2\arcmin.}}
     \label{Fig_coldens_compare}%
    \end{figure*}
}

\end{document}